# Study of the Behavior of the Nuclear Modification Factor in Freeze-out State

M. Ajaz∗,[1], M. K. Suleymanov[1],[2], E. U. Khan[1], M. Q. Haseeb[1], K. H. Khan[1], Z. Wazir1

[1] Department of Physics
COMSATS Institute of Information Technology
44000 Islamabad, Pakistan

[2] Veksler - Baldin Laboratory of High Energy
Joint Institute for Nuclear Research
141980, Dubna, Russia

∗ rph_f06_013@comsats.edu.pk

## 1 Introduction

One of the latest trends in the advancement of experimental high-energy physics is to identify the Quark Gluon Plasma (QGP) [1] predicted qualitatively by the Quantum Chromodynamics (QCD) [2]. QGP is considered to be a state of strongly interacting matter under extreme conditions (high temperatures and/or densities of the baryons).This can be brought about under laboratory conditions during collisions of relativistic heavy nuclei by increasing energies and varying masses of colliding nuclei. This leads to a continuing quest of leading research centers on high-energy physics to create new accelerators of heavy nuclei and enhance the energies of existing accelerators. Strongly interacting matter may be subject to a series of phase transitions with increasing temperature and/or density of the baryon among which is the first-order phase transition of restoration of a specific symmetry of strong interactions—chiral symmetry that is strongly violated at low temperatures and/or densities of the baryon charge. However, to create necessary laboratory conditions and pick up a "signal" of formation of the QGP phase, one needs a lot of intellectual and material resources. The well known time evolution of central heavy ion collisions is shown in Fig. 1. Here the five states are shown as: I-pre-interaction state; II-parton-parton interaction one; III-mixed phase; IV-QGP phase ; V- freezout. In each state the matter can be characterized by different temperature and density. Apart from these parameters, there could be another very interesting parameter, namely the transparency (T r ) of matter, to characterize these states. We believe that the appearance and changing of the transparency could give the necessary information for identification the QGP formation.

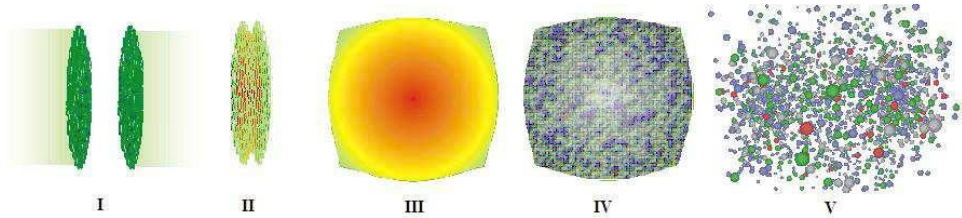

Figure 1: Time evolution of the central heavy ion collisions.

In Ref. [3] a signal on mixed phase formation is considered as an important point to identify QGP because such a state mast contain real particles which can interact with matter and therefore could be detected.

## 2   Nuclear Modification Factor

In this paper we discuss some ideas connected with identification of the QGP using the information coming from the freezout state. The main idea is that the values of transparency (T r) for different states of time evolution of heavy ion collisions are different (T $r_{III}$, T $r_{IV}$ and T $r_V$).

To characterize the  T r it is convenient to use the nuclear modification factor (R). A comparison of yields in different ion systems by using nuclear modification factors (involving Central and Peripheral collisions) should provide information on the hadronization [5] (see Fig. 2. R highlights the particle type dependence at intermediate $p_T$ as was suggested by coalescence models [6] leading to the idea that hadrons result from the coalescence of quarks in the dense medium. Fig. 3 shows the recent results from RHIC on heavy flavor production [7].

It is supposed that T $r_{III-V}$   ∼ $R_{III-V}$ . To restore the time scale we are going to use
the values of temperature because they must be different for III-V states. If the states of III-V will appear critically so the regime change have to be observed in the behavior of R as a function of temperature.

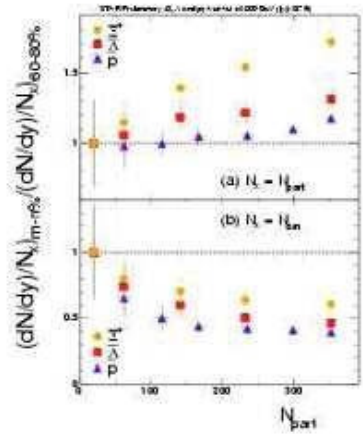

Figure 2: Yields of some baryons as a function of the centrality (expressed with the number of participants) in Au+Au collisions normalized to the most peripheral point and to Npart (upper frame) and $N_{bin}$(lower frame).

## 3  Results

To confirm the above idea we use data coming from different heavy ion generators and experiments. Fig. 4 shows the result of the study of the behavior of R function defined as a ratio of the yields of different particles at central to peripheral collision as a function of the thermal freeze out temperature ($T_{th}$) produced in Au-Au collisions at RHIC energies. The data is simulated using the Fast Hadron Freezout Generater (FASTMC) [4]. The FASTMC hadron generation allows one to study and analyze various observables for stable hadrons and hadron resonances produced in ultra-relativistic heavy ion collisions. Particles can be generated on the chemical orthermal freeze-out hyper surface represented by a parametrization or a numerical solution of relativistic hydrodynamics with given initial conditions and equation of state [4]. There are two regions in the behavior of R as a function of the $T_{th}$(see Fig. 4). In the first region one can see that in the freeze out state R is almost a linearly increasing function of the $T_{th}$ independent of the types of particles and the second region is a straight line, which has no dependence on $T_{th}$, it can be regarded as a regime change.

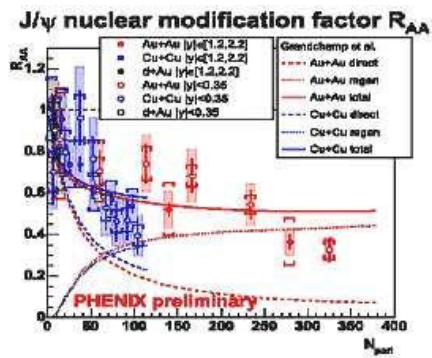

Figure 3: Nuclear modification factor as function of a number of participant.

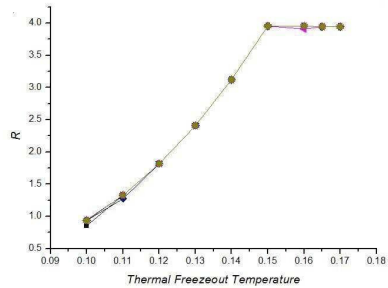

Figure 4: Nuclear Modification Factor as a function of thermal freezeout temperature.

References

[1] S. Jeon and V. Koch Review for Quark-Gluon Plasma 3 eds. R. C. Hwa and X.-N. Wang (World Scientific, Singapore, 2004.

[2] Walter Greiner, Andreas Schafer, (1994). Quantum Chromodynamics. Springer. ISBN 0-387-57103-5.

[3] A. N. Sissakian, A. S. Sorin, M. K. Suleymanov, V. D. Toneev, G. M. Zinovjev. Phys.Part.Nucl.Lett.5, 1-6 (2008)

[4] N. S. Amelin et. al., arXiv:0711.0835 (2007)

[5] Christelle Roy for the STAR Collaboration, POS (HEP2005) 141.

[6] D. Molnar et al., Phys. Rev. Lett. 91, 092301 (2003)

[7] A. A. P. Suaide. Brazilian Journal of Physics 37 N 2C, 731-735 (2007)